\documentclass[twocolumn,preprintnumbers,amsmath,aps]{revtex4}
\usepackage{graphicx}
\usepackage{dcolumn}
\usepackage{bm}
\usepackage{color}
\usepackage{multirow}
\begin{document}
\def\eq#1{(\ref{#1})}
\def\fig#1{Fig.\hspace{1mm}\ref{#1}}
\def\tab#1{Tab.\hspace{1mm}\ref{#1}}
\title{High-pressure superconducting state in hydrogen}
\author{A. M. Duda$^{\left(1\right)}$}
\email{aduda@wip.pcz.pl}
\author{R. Szcz{\c{e}}{\'s}niak$^{\left(1, 2\right)}$}
\email{szczesni@wip.pcz.pl} 
\author{M. A. Sowi{\'n}ska$^{\left(1\right)}$}
\author{A. H. Kosiacka$^{\left(1\right)}$}
\affiliation{$^1$ Institute of Physics, Cz{\c{e}}stochowa University of Technology, Ave. Armii Krajowej 19, 42-200 Cz{\c{e}}stochowa, Poland}
\affiliation{$^2$ Institute of Physics, Jan D{\l}ugosz University in Cz{\c{e}}stochowa, Ave. Armii Krajowej 13/15, 42-200 Cz{\c{e}}stochowa, Poland}
\date{\today} 
\begin{abstract}
The paper determines the thermodynamic parameters of the superconducting state in the metallic atomic hydrogen under the pressure at $1$ TPa, $1.5$ TPa, and $2.5$ TPa. The calculations were conducted in the framework of the Eliashberg formalism. It has been shown that the critical temperature is very high (in the range from $301.2$ K to $437.3$ K), as well as high are the values of the electron effective mass 
(from $3.43$ $m_{e}$ to $6.88$ $m_{e}$), where $m_{e}$ denotes the electron band mass. The ratio of the low-temperature energy gap to the critical temperature explicitly violates the predictions of the BCS theory: $2\Delta\left(0\right)/k_{B}T_{C}\in\left<4.84,5.85\right>$. Additionally, the free energy difference between the superconducting and normal state, the thermodynamic critical field, and the specific heat of the superconducting state have been determined. Due to the significant strong-coupling and retardation effects those quantities cannot be correctly described in the framework of the BCS theory.
\end{abstract}\ 

\maketitle

\section{Introduction}

The superconducting state with the possibly high value of the critical temperature ($T_{C}$) is one of the most important goals of the solid state physics.

Initially the greatest hopes were related to the group of the superconductors discovered in 1986 by Bednorz and M{\"u}ller 
(the so-called cuprates) \cite{Bednorz1986A}, \cite{Bednorz1988A}. Unfortunately, the years of study within the family of the compounds under consideration allowed to obtain the maximum value of $T_{C}$ equal only to $135$ K (${\rm HgBa_{2}Ca_{2}Cu_{3}O_{8+y}}$) \cite{Chu1993A}. 
However, the critical temperature could still be slightly increased up to $T_{C}=164$ K, when increasing the external pressure ($p$) up to the value of about $31$ GPa \cite{Gao1994A}, or up to $153$ K for $p=15$ GPa, which is suggested in the paper \cite{Takeshita2013A}.

Let us notice that the new families of the superconductors discovered in the later years (the fulleride, the iron-based, and the ${\rm MgB_{2}}$-based compounds \cite{Hebard1991A}, \cite{Kamihara2008A}, \cite{Nagamatsu2001A}) have been characterized by the significantly lower values of the critical temperature than cuprates.
 
Alongside the mainstream of the research, also the search for the high-temperature superconducting state in the more exotic physical systems was conducted. The most promising direction is connected with the superconducting state inducing in the metallic hydrogen (Ashcroft in 1968 \cite{Ashcroft1968A}). 

The predicted high value of the critical temperature for the superconducting state in hydrogen is related to the following facts: 
(i) the large value of the Debye frequency resulting from the small mass of the atomic nucleus (single proton) and 
(ii) lack of the electrons on the inner shells, which should result in the strong coupling of the electron-phonon type \cite{Maksimov2001A}, \cite{Szczesniak2009A}.   

Unfortunately, the theoretical predictions has not been able to be confirmed experimentally to the present day, which results from the very high value of the pressure of hydrogen’s metallization ($p_{m}\sim 400$ GPa) \cite{Stadele2000A}. However, recent experimental data obtained for the compounds ${\rm H_{2}S}$ and ${\rm H_{3}S}$ ($\left[T_{C}\right]_{\rm max}\sim 200$ K \cite{Drozdov2014A}, \cite{Drozdov2015A}), where the chemical pre-compression lowers the value of $p_{m}$ \cite{Ashcroft2004A}, indirectly confirms the results of the theoretical considerations for hydrogen \cite{Li2014A}, \cite{Duan2014A}, \cite{Durajski2015A}, and \cite{Durajski2015B}. 

Referring specifically to the theoretical results obtained for the superconducting state in hydrogen, the attention has to be paid to the fact that the value of $T_{C}$ is high in the whole range of the pressure from about $400$ GPa to $3.5$ TPa (the pressure near the core of the planet of the Jovian-type \cite{Guillot2004A}). In particular, for the molecular phase of hydrogen ($p\in\left<400,500\right>$ GPa), the critical temperature grows rapidly from about $80$ K to $350$ K \cite{Cudazzo2008A}, \cite{Szczesniak2013A}, \cite{Yan2011A}, and \cite{Szczesniak2012A}. Above $500$ GPa, the value of $T_{C}$ stabilizes in the range from $\sim 300$ K to $\sim 470$ K, whereas for $2$ TPa, the maximum of the critical temperature – able to reach even the value of $630$ K is predicted \cite{Maksimov2001A}, \cite{Szczesniak2009A}. 

The thermodynamics of the superconducting state in hydrogen has been studied for the few selected pressures \cite{Szczesniak2009A}, \cite{Szczesniak2013A}, \cite{Szczesniak2012A}. The obtained results suggest that, due to the significant strong-coupling and retardation effects, the description with the use of the BCS theory \cite{Bardeen1957A}, \cite{Bardeen1957B} is not sufficient and the Eliashberg method should be used instead \cite{Eliashberg1960A}.  

The thermodynamic parameters of the superconducting state in hydrogen for the pressure at $1$ TPa, $1.5$ TPa, and $2.5$ TPa have been determined in the present work. The calculated values of the coupling constant ($\lambda$) and the logarithmic frequency ($\omega_{\rm ln}$) in the considered case prove that the parameter determining the magnitude of the strong-coupling and retardation effects ($r=k_{B}T_{C}/\omega_{\rm ln}$) significantly deviates from the limit value of BCS: $\left[r\right]_{\rm BCS}=0$ (see \tab{t1}). For this reason, the thermodynamics of the superconducting state was described with the help of the Eliashberg equations \cite{Eliashberg1960A}.    

\section{The formalism}

The Eliashberg equations on the imaginary axis ($i=\sqrt{-1}$) take the following form: 
\begin{equation}
\label{r1}
\phi_{m}=\frac{\pi}{\beta}\sum_{n=-M}^{M}
\frac{\lambda\left(i\omega_{m}-i\omega_{n}\right)-\mu^{\star}\theta\left(\omega_{c}-|\omega_{n}|\right)}
{\sqrt{\omega_n^2Z^{2}_{n}+\phi^{2}_{n}}}\phi_{n},
\end{equation}
\begin{equation}
\label{r2}
Z_{m}=1+\frac{1}{\omega_{m}}\frac{\pi}{\beta}\sum_{n=-M}^{M}
\frac{\lambda\left(i\omega_{m}-i\omega_{n}\right)}{\sqrt{\omega_n^2Z^{2}_{n}+\phi^{2}_{n}}}\omega_{n}Z_{n}.
\end{equation}
The order parameter is defined by the ratio: $\Delta_{m}=\phi_{m}/Z_{m}$, where $\phi_{m}=\phi\left(i\omega_{m}\right)$ represents the order parameter function and $Z_{m}=Z\left(i\omega_{m}\right)$ is the wave function renormalization factor. The $m$-th Matsubara frequency is given by: $\omega_{m}=\left(\pi /\beta\right)\left(2m-1\right)$, where: $\beta=\left(k_{B}T\right)^{-1}$. The pairing kernel is given with the following formula: $\lambda\left(z\right)= 2\int_0^{\Omega_{\rm{max}}}d\Omega\frac{\Omega}{\Omega ^2-z^{2}}\alpha^{2}F\left(\Omega\right)$, 
where $\alpha^{2}F\left(\Omega\right)$ is the Eliashberg function. The Eliashberg functions were calculated in the paper \cite{McMahon2011A} for the cases under consideration. The values of the maximum phonon frequency ($\Omega_{\rm{max}}$) are collected in \tab{t1}.

The depairing correlations were modelled parametrically with the help of the Coulomb pseudopotential: 
$\mu^{\star}\in\left\{0.1, 0.2, 0.3\right\}$. $\theta$ denotes the Heaviside function, $\omega_{c}$ represents the cut-off frequency: 
$\omega_{c}=5\Omega_{\rm{max}}$. 

The Eliashberg equations were solved for $M=1100$, which ensured the stability of the functions $\phi_{m}$ and $Z_{m}$ for the temperatures larger than, or equal to $T_{0}=50$ K. The numerical modules described and tested in the papers: \cite{Szczesniak2013A}, \cite{Szczesniak2012A}, \cite{Szczesniak2013B}, \cite{Drzazga2014A}, \cite{Szczesniak2014A}, \cite{Szczesniak2014B}, and \cite{Szczesniak2015A} were used.

In order to accurately determine the value of the energy gap and the electron effective mass, the solutions of the Eliashberg equations from the imaginary axis should be analytically extended on the real axis ($\phi_{m}\rightarrow\phi\left(\omega\right)$ and 
$Z_{m}\rightarrow Z\left(\omega\right)$). The following equations were used for this purpose:
\begin{widetext}
\begin{eqnarray}
\label{r3}
\phi\left(\omega+i\delta\right)&=&
                                  \frac{\pi}{\beta}\sum_{m=-M}^{M}
                                  \left[\lambda\left(\omega-i\omega_{m}\right)-\mu^{\star}\theta\left(\omega_{c}-|\omega_{m}|\right)\right]
                                  \frac{\phi_{m}}
                                  {\sqrt{\omega_m^2Z^{2}_{m}+\phi^{2}_{m}}}\\ \nonumber
                              &+& i\pi\int_{0}^{+\infty}d\omega^{'}\alpha^{2}F\left(\omega^{'}\right)
                                  \left[\left[N\left(\omega^{'}\right)+f\left(\omega^{'}-\omega\right)\right]
                                  \frac{\phi\left(\omega-\omega^{'}+i\delta\right)}
                                  {\sqrt{\left(\omega-\omega^{'}\right)^{2}Z^{2}\left(\omega-\omega^{'}+i\delta\right)
                                  -\phi^{2}\left(\omega-\omega^{'}+i\delta\right)}}\right]\\ \nonumber
                              &+& i\pi\int_{0}^{+\infty}d\omega^{'}\alpha^{2}F\left(\omega^{'}\right)
                                  \left[\left[N\left(\omega^{'}\right)+f\left(\omega^{'}+\omega\right)\right]
                                  \frac{\phi\left(\omega+\omega^{'}+i\delta\right)}
                                  {\sqrt{\left(\omega+\omega^{'}\right)^{2}Z^{2}\left(\omega+\omega^{'}+i\delta\right)
                                  -\phi^{2}\left(\omega+\omega^{'}+i\delta\right)}}\right],
\end{eqnarray}
and
\begin{eqnarray}
\label{r4}
Z\left(\omega+i\delta\right)&=&
                                  1+\frac{i}{\omega}\frac{\pi}{\beta}\sum_{m=-M}^{M}
                                  \lambda\left(\omega-i\omega_{m}\right)
                                  \frac{\omega_{m}Z_{m}}
                                  {\sqrt{\omega_m^2Z^{2}_{m}+\phi^{2}_{m}}}\\ \nonumber
                              &+&\frac{i\pi}{\omega}\int_{0}^{+\infty}d\omega^{'}\alpha^{2}F\left(\omega^{'}\right)
                                  \left[\left[N\left(\omega^{'}\right)+f\left(\omega^{'}-\omega\right)\right]
                                  \frac{\left(\omega-\omega^{'}\right)Z\left(\omega-\omega^{'}+i\delta\right)}
                                  {\sqrt{\left(\omega-\omega^{'}\right)^{2}Z^{2}\left(\omega-\omega^{'}+i\delta\right)
                                  -\phi^{2}\left(\omega-\omega^{'}+i\delta\right)}}\right]\\ \nonumber
                              &+&\frac{i\pi}{\omega}\int_{0}^{+\infty}d\omega^{'}\alpha^{2}F\left(\omega^{'}\right)
                                  \left[\left[N\left(\omega^{'}\right)+f\left(\omega^{'}+\omega\right)\right]
                                  \frac{\left(\omega+\omega^{'}\right)Z\left(\omega+\omega^{'}+i\delta\right)}
                                  {\sqrt{\left(\omega+\omega^{'}\right)^{2}Z^{2}\left(\omega+\omega^{'}+i\delta\right)
                                  -\phi^{2}\left(\omega+\omega^{'}+i\delta\right)}}\right]. 
\end{eqnarray}
\end{widetext}
The symbols $N\left(\omega\right)$ and $f\left(\omega\right)$ are the Bose-Einstein and the Fermi-Dirac functions, respectively. 

\subsection{THE OBTAINED RESULTS}
\begin{table}
\caption{\label{t1} The selected parameters of the high-pressure superconducting state in hydrogen.}
\begin{center}
\begin{ruledtabular}
\begin{tabular}{c|c|c|c|c|c}
                        &            &                     &               &                    &                          \\
      Quantity          & Unit\hspace{3mm}& $\mu^{\star}$       & $1$ TPa       & $1.5$ TPa          & $2.5$ TPa           \\
                        &            &                     &               &                    &                          \\
\hline
                        &            &                     &               &                    &                          \\
$\lambda$               &            &                     & 5.88          & 4.71               & 2.43                     \\
                        &            &                     &               &                    &                          \\
$\omega_{\rm ln}$       & meV        &                     & 50.1          & 43.18              & 147.22                   \\
                        &            &                     &               &                    &                          \\
                        &            & 0.1                 & 0.802         & 0.827              & 0.256                    \\
$r$                     &            & 0.2                 & 0.679         & 0.690              & 0.210                    \\
                        &            & 0.3                 & 0.606         & 0.601              & 0.180                    \\
                        &            &                     &               &                    &                          \\
\hline                        
                        &            &                     &               &                    &                          \\
$\Omega_{\rm max}$      & meV        &                     & 479.40        & 544.55             & 650.47                   \\
                        &            &                     &               &                    &                          \\                        
\hline
                        &            &                     &               &                    &                          \\
                        &            & 0.1                 & 466.1         & 414.1              & 437.3                    \\
$T_{C}$                 & K          & 0.2                 & 395.1         & 345.7              & 359.3                    \\
                        &            & 0.3                 & 352.1         & 301.2              & 308.0                    \\
                        &            &                     &               &                    &                          \\
                        &            & 0.1                 & 106.06        & 89.03              & 89.80                    \\
$\Delta\left(0\right)$  & meV        & 0.2                 & 89.73         & 73.42              & 72.37                    \\
                        &            & 0.3                 & 78.93         & 63.37              & 61.24                    \\
                        &            &                     &               &                    &                          \\                                                
\hline 
                        &            &                     &               &                    &                          \\
                        &            & 0.1                 & 612.16        & 489.6              & 476.7                    \\
$H_{C}\left(0\right)/\sqrt{\rho\left(0\right)}$& meV & 0.2 & 533.98        & 416.8              & 392.6                    \\
                        &            & 0.3                 & 480.06        & 364.0              & 336.8                    \\
                        &            &                     &               &                    &                          \\
                        &            & 0.1                 & 2871.3        & 2065.35            & 1879.4                   \\
$\Delta C\left(T_{C}\right)/k_{B}\rho\left(0\right)$& meV & 0.2 & 2677.67  & 1584.87            & 1505.9                   \\
                        &            & 0.3                 & 3265.82       & 1652.47            & 1262.8                   \\
                        &            &                     &               &                    &                          \\
\end{tabular}
\end{ruledtabular}
\end{center}
\end{table}
\begin{figure*}
\centering
\includegraphics[width=1.6\columnwidth]{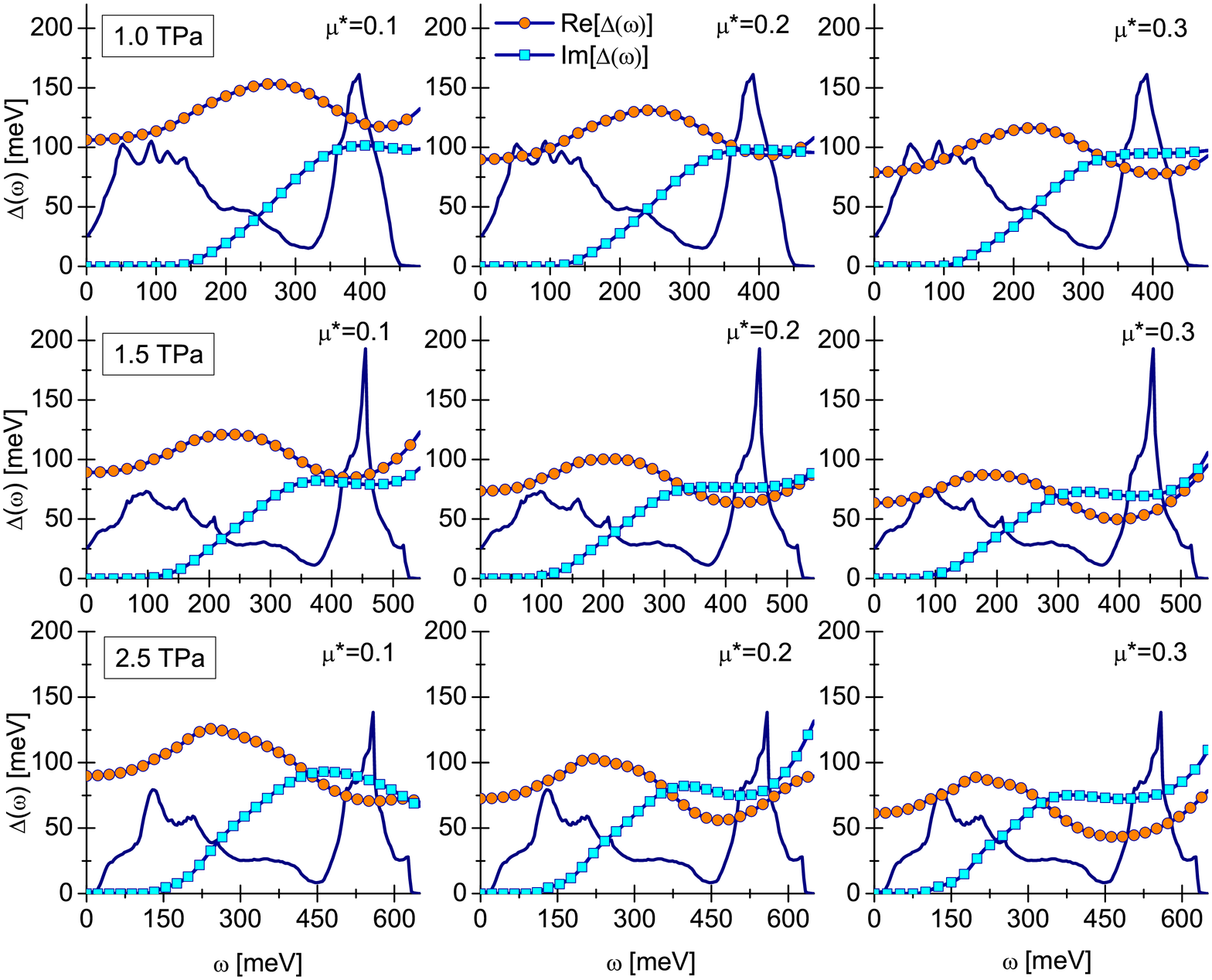}
\caption{The real part and the imaginary part of the order parameter on the real axis for $T=T_{0}$. Additionally, the rescaled Eliashberg functions are plotted (100$\alpha^{2}F\left(\omega\right)$), which reflects the correlations in the course of the functions 
${\rm Re\left[\Delta\left(\omega\right)\right]}$ and $\alpha^{2}F\left(\omega\right)$.}
\label{f1}
\includegraphics[width=1.9\columnwidth]{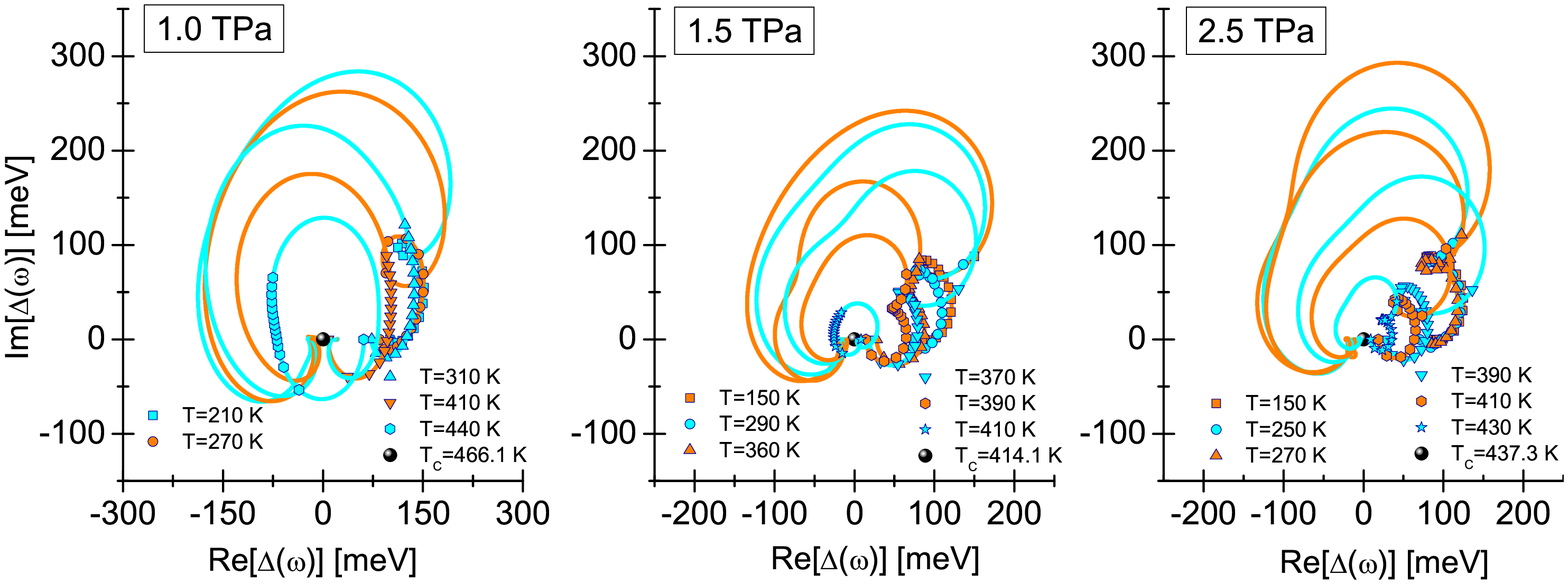}
\caption{The real part and the imaginary part of the order parameter on the complex plane for the selected values of the temperature and 
$\mu^{\star}=0.1$. The lines with symbols were obtained for $\omega\in\left<0,\Omega_{\rm max}\right>$, the lines without symbols were obtained for $\omega\in\left(0,\omega_{c}\right>$.}
\label{f2}
\end{figure*}
\fig{f1} presents the form of the order parameter on the real axis for the lowest temperature analysed in the paper. It can be easily noticed that the non-zero values are taken only by the real part of the order parameter in the range of the low frequencies. Physically it means no damping effects, which is tantamount to the forever living Cooper pairs \cite{Varelogiannis1997A}. Additionally, clearly noticeable is the destructive impact of the increase in the value of the Coulomb pseudopotential on the superconducting state.      

The open dependence of the order parameter on the temperature is depicted in \fig{f2}. It has been found that the values of the function 
$\Delta\left(\omega\right)$ on the complex plane construct the characteristic deformed spirals, whose size clearly decreases with the increasing temperature.       

The physical value of the order parameter for the given temperature has been calculated on the basis of the formula:
\begin{equation}
\label{r5}
\Delta\left(T\right)={\rm Re}\left[\Delta\left(\omega=\Delta\left(T\right)\right)\right].
\end{equation}

\begin{figure*}
\centering
\includegraphics[width=1.9\columnwidth]{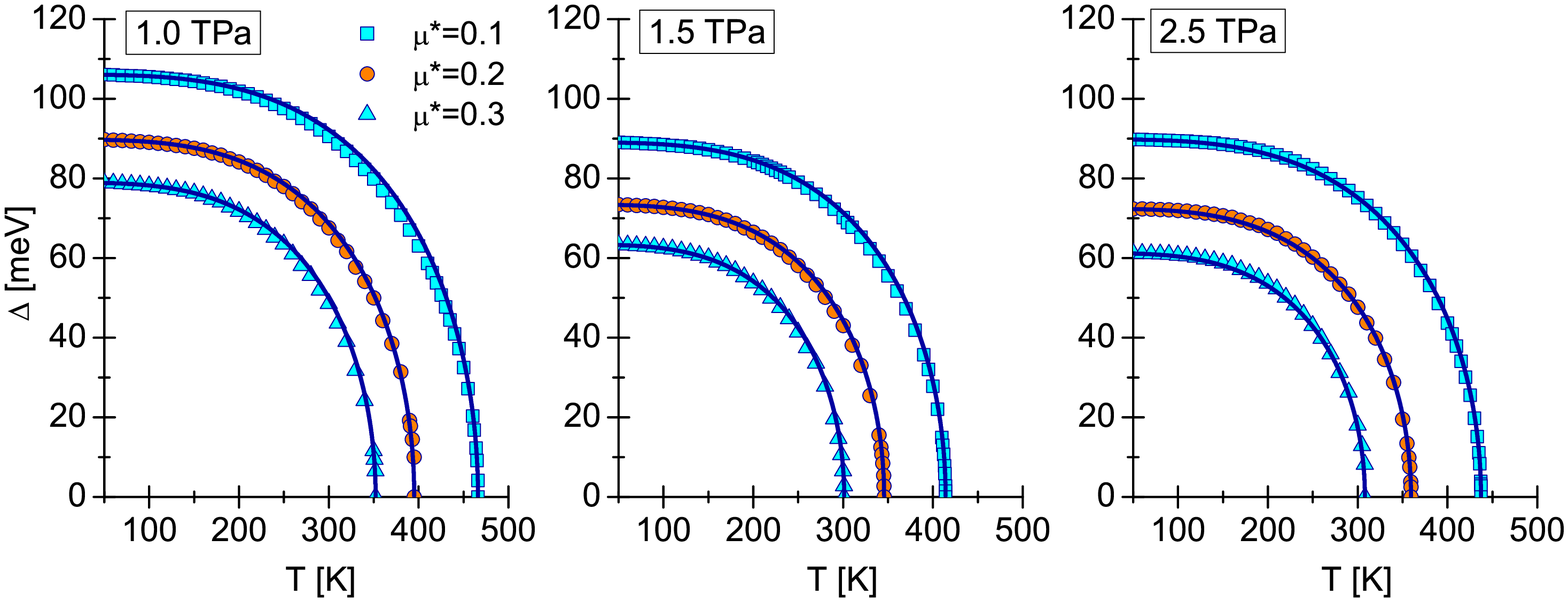}
\caption{The dependence of the order parameter on the temperature.}
\label{f3}
\includegraphics[width=1.9\columnwidth]{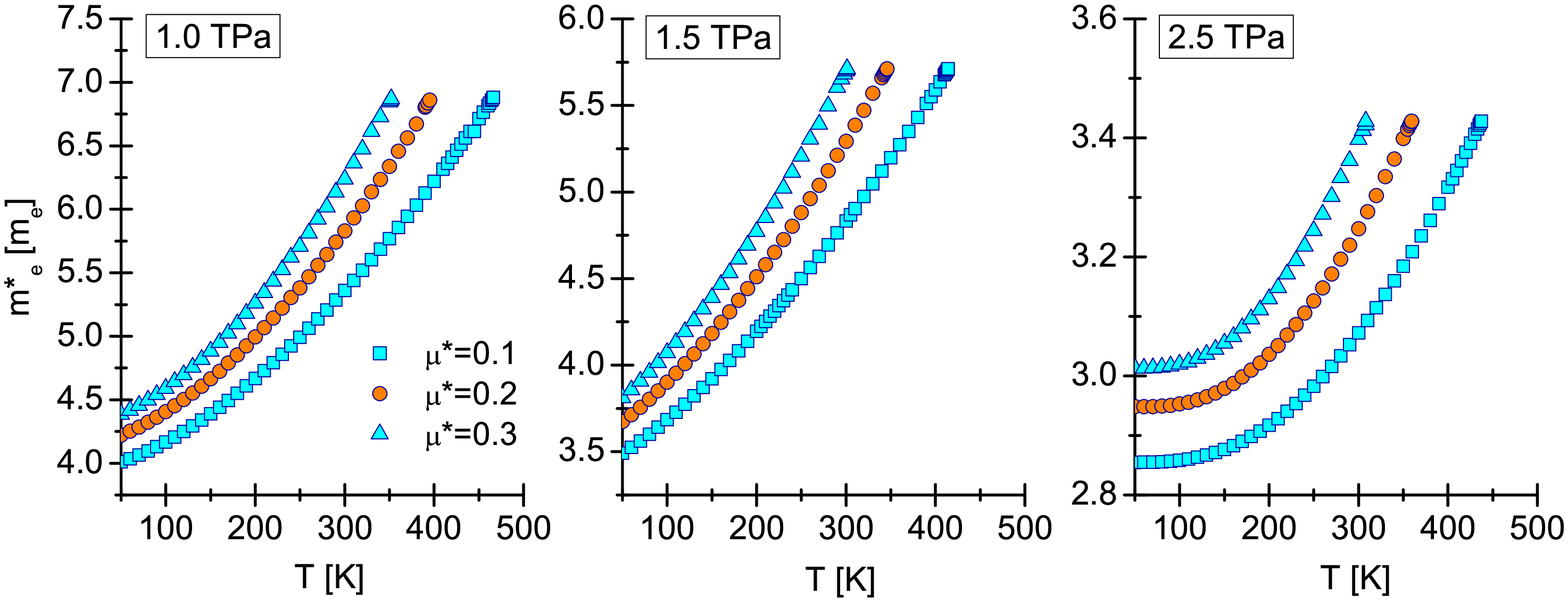}
\caption{The influence of the temperature on the value of the electron effective mass.}
\label{f4}
\end{figure*}

The obtained results are plotted in \fig{f3}. It can be clearly seen that, irrespective of the assumed magnitude of the depairing electron correlations, the order parameter for $T=T_{0}$ and the critical temperature take the high values. In particular, $T_{C}$ changes in the range from about $300$ K to $470$ K, whereas $\Delta\left(0\right)=\Delta\left(T_{0}\right)$ lies in the range from about $61$ meV to $106$ meV 
(see also \tab{t1}). Let us also notice that due to the strong-coupling and retardation effects, the ratio of the energy gap to the critical temperature clearly exceeds the universal value of $3.53$, which is predicted by the BCS theory \cite{Bardeen1957A}, \cite{Bardeen1957B}: 
$2\Delta\left(0\right)/k_{B}T_{C}\in\left<4.84,5.85\right>$. 
The full dependence of the order parameter on the temperature can be reproduced with the help of the simple formula:
\begin{equation}
\label{r6}
\Delta\left(T,\mu^{\star}\right)=\Delta\left(\mu^{\star}\right)\sqrt{1-\left(\frac{T}{T_{C}}\right)^{\Gamma}}, 
\end{equation}
where $\Gamma=3.2$.
The additional consequence of the extremely strong electron-phonon interaction in the metallic atomic hydrogen is the significant increase in the value of the electron effective mass: $m^{\star}_{e}={\rm Re}\left[Z\left(\omega=0\right)\right]m_{e}$, where the symbol $m_{e}$ represents the electron band mass. 
\begin{figure*}
\centering
\includegraphics[width=1.9\columnwidth]{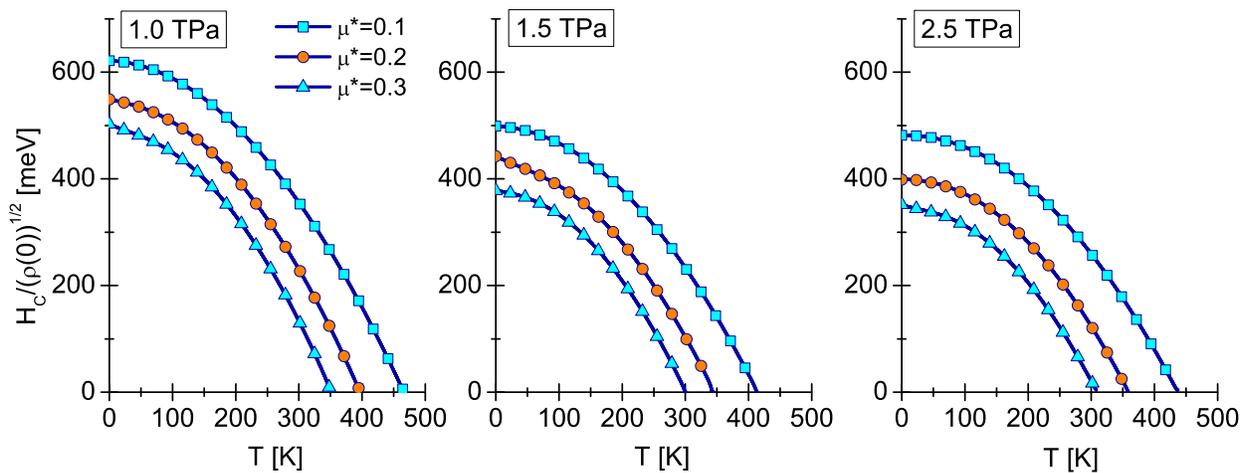}
\caption{The thermodynamic critical field as a function of the temperature.}
\label{f5}
\end{figure*}
\begin{figure*}
\centering
\includegraphics[width=1.9\columnwidth]{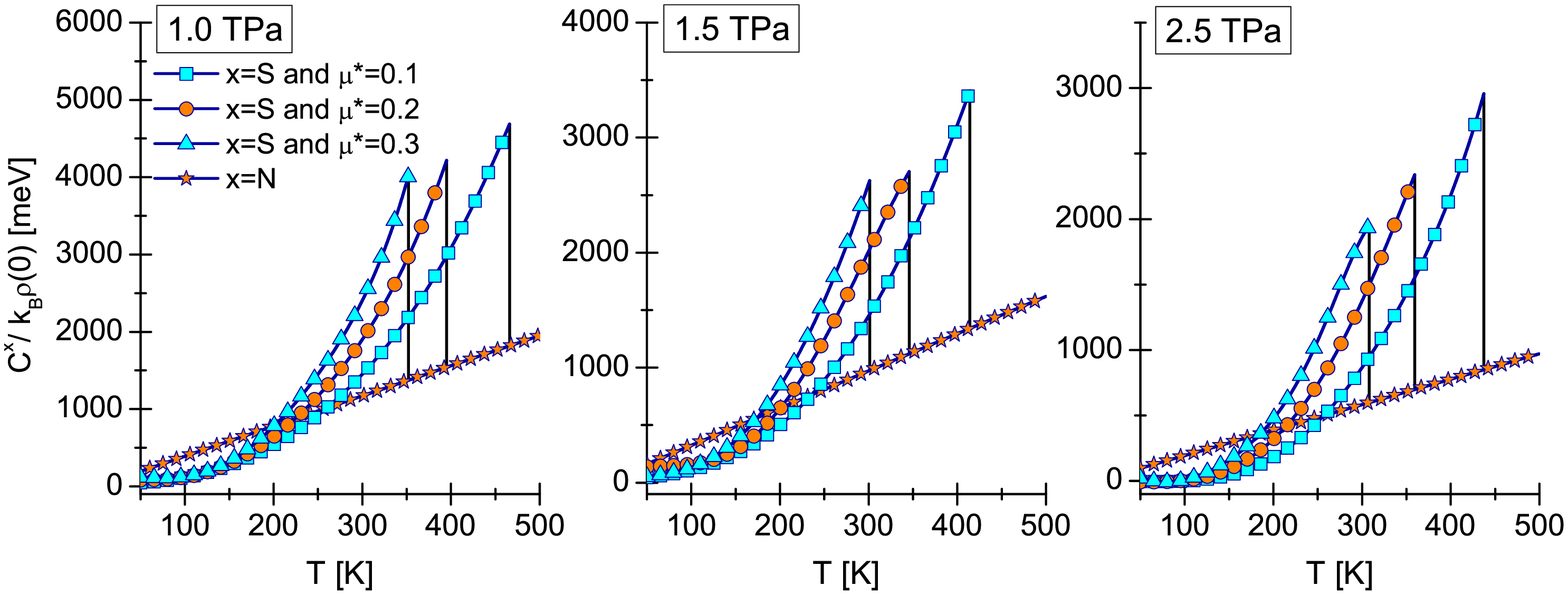}
\caption{The dependence of the specific heat of the superconducting state and the normal state on the temperature.}
\label{f6}
\end{figure*}
The detailed course of the function $m^{\star}_{e}\left(T\right)$ is presented in \fig{f4}. It can be seen that the effective mass takes the particularly high values for the pressure at $1$ TPa and $1.5$ TPa (the range from $3.49$ $m_{e}$ to $6.88$ $m_{e}$). However, for $p=2.5$ TPa the values of $m^{\star}_{e}\left(T\right)$ are also significant (the range from $2.85$ $m_{e}$ to $3.43$ $m_{e}$). The maximums of the plotted functions $m^{\star}_{e}\left(T\right)$ are always observed for $T=T_{C}$, where: 
$\left[m^{\star}_{e}\right]_{\rm max}\simeq\left(1+\lambda\right)m_{e}$ \cite{Carbotte1990A}.  
The free energy difference between the superconducting and normal state has been calculated on the basis of the solutions of the Eliashberg equations on the imaginary axis \cite{Carbotte1990A}:  
\begin{eqnarray}
\label{r7}
\frac{\Delta F}{\rho\left(0\right)}&=&-\frac{2\pi}{\beta}\sum_{m=1}^{M}
\left(\sqrt{\omega^{2}_{m}+\Delta^{2}_{m}}- \left|\omega_{m}\right|\right)\\ \nonumber
&\times&(Z^{S}_{m}-Z^{N}_{m}\frac{\left|\omega_{m}\right|}
{\sqrt{\omega^{2}_{m}+\Delta^{2}_{m}}}),
\end{eqnarray}  
where $Z^{S}_{m}$ and $Z^{N}_{m}$ denote the wave function renormalization factor for the superconducting state ($S$) and the normal state ($N$), respectively. The symbol $\rho\left(0\right)$ denotes the value of the electron density of states at the Fermi level. Hence, the thermodynamic critical field is determined in the very simple manner:
\begin{equation}
\label{r8}
\frac{H_{C}}{\sqrt{\rho\left(0\right)}}=\sqrt{-8\pi\left[\Delta F/\rho\left(0\right)\right]},
\end{equation}
as well as the specific heat difference between the superconducting and normal state:
\begin{equation}
\label{r9}
\frac{\Delta C\left(T\right)}{k_{B}\rho\left(0\right)}=-\frac{1}{\beta}\frac{d^{2}\left[\Delta F/\rho\left(0\right)\right]}{d\left(k_{B}T\right)^{2}},
\end{equation}
where the specific heat of the normal state is given with the expression: $C^{N}=\gamma{T}$, $\gamma$ denotes the Sommerfeld constant: 
$\gamma=\frac{2}{3}\pi^{2}k_{B}^{2}\rho(0)\left(1+\lambda\right)$.

The obtained results are presented in \fig{f5} and \fig{f6}. Using the plots, it can be seen that the characteristic values of the analysed thermodynamic functions (\tab{t1}) decrease with the increasing pressure.

\section{Summary}

The superconducting state inducing in the metallic atomic hydrogen for the value of the pressure at $1$ TPa, $1.5$ TPa, and $2.5$ TPa cannot be properly characterized by the BCS theory. This is due to the existence of the significant strong-coupling and retardation effects. In particular, they are responsible for the very high values of the critical temperature (the range from about $300$ K to $470$ K) and the marked increase in the effective mass of the electron. It should be noted that the results under consideration were obtained for the wide range of values of the depairing electron correlations: $\mu^{\star}\in\left\{0.1, 0.2, 0.3\right\}$. Additionally, the presented work has determined the dependence of the thermodynamic critical field and the specific heat of the superconducting state on the temperature. The characteristic values of the discussed functions clearly decrease with the increasing pressure.

\bibliography{template}
\end{document}